\documentstyle[12pt]{article}
\def\be{\begin{equation}}
\def\ee{\end{equation}}
\def\ba{\begin{array}{c}}
\def\baa{\begin{array}{cc}}
\def\ea{\end{array}}

\def\ben{$$}
\def\een{$$}
\begin{document}

%\titlepage
%\vspace*{2cm}

\begin{center}{\Large \bf
The method of Hill determinants in PT-symmetric quantum mechanics
 }\end{center}

\vspace{5mm}

\begin{center}
Miloslav Znojil
\vspace{3mm}

\'{U}stav jadern\'e fyziky AV \v{C}R, 250 68 \v{R}e\v{z},
Czech Republic\\

\end{center}

\vspace{5mm}

\section*{Abstract}
Hill-determinant method is described and shown applicable within
the so called PT-symmetric quantum mechanics. We demonstrate that
in a way paralleling its traditional Hermitian applications and
proofs the method guarantees the necessary asymptotic decrease of
wave functions $\psi(x)$ as resulting from a fine-tuned mutual
cancellation of their asymptotically growing exponential
components. Technically, the rigorous proof is needed/offered that
in a quasi-variational spirit the method allows us to work, in its
numerical implementations, with a sequence of truncated forms of
the rigorous Hill-determinant power series for the normalizable
bound states $\psi(x)$.

\vspace{9mm}

\noindent
 PACS 03.65.Ge, 03.65.Fd

%\vspace{9mm}

%\begin{center}
%{\small \today, jAHOr.TEX }
%\end{center}

% \newpage

\section{Introduction}

One of the key sources of an enormous popularity of
one-dimensional anharmonic-oscillator Schr\"{o}dinger equations
 \be
  \left
(-\,\frac{d^2}{dx^2} + V(x)
 \right )
 \, \psi(x) =
E  \, \psi(x), \ \ \ \ \ x \in (-\infty,\infty)
 \label{SEa}
 \ee
with polynomial potentials, say, of the quartic form
 \be
V(x)=a\,x^4+b\,x^3+c\,x^2+d\,x, \ \ \ \ \ \ \ a = 1 \label{AHO}
 \ee
lies, definitely, in their methodically illuminating role of
apparently ``next-to-solvable" examples in quantum theory.

Thirty five years ago one of the characteristic results in this
direction has been obtained by Bender and Wu \cite{BW} who made a
highly nontrivial observation that in some models of such a type,
{\em all} the pertaining bound-state energies $E_n$ may be
understood as values of a {\em single} analytic function when
evaluated on its different Riemann sheets. This observation (made
at $b=d=0$) was a real breakthrough in the intensive contemporary
perturbative analysis of eq. (\ref{SEa}) + (\ref{AHO}). Later on,
it has been complemented by the alternative semi-classical
\cite{Voros} and quasi-variational \cite{Hill} sophisticated
treatments of eq. (\ref{AHO}). In this sense, the model provided
an extremely useful insight in the possible structures encountered
within relativistic quantum field theory \cite{Simon}.

The hypotheses of ref. \cite{BW} have also been explicitly
verified in the phenomenologically equally important
cubic-oscillator limit $a\to 0$ \cite{Alvarez}. In parallel, the
closely interrelated semi-classical and quasi-variational (often
called Hill-determinant, HD) constructions of bound states were
subsequently generalized to virtually all polynomials $V(x)$. In
the former semi-classical setting, an updated sample of references
is offered by these proceedings. In contrast, the parallel HD
approach to polynomial $V(x)$ \cite{prd} seems ``forgotten" at
present, in spite of its extremely useful capability of
complementing the perturbative and semi-classical constructions.
This is one of the reasons why the forthcoming text has been
written for the same proceedings.

A nice illustration of the HD -- semi-classical complementarity of
the corresponding algorithms has been thoroughly described in ref.
\cite{riccatipade} (here, we shall skip this illustration
referring to the original paper containing a few numerical tables
and further references).

The HD method emphasizes certain important non-perturbative
features of the energies. Again we would like to recollect an
elementary illustrative two-dimensional example of ref.
\cite{exotic}, or a more extensive one-dimensional
sextic-oscillator study \cite{Singh} where the HD-type energies
have been specified as roots of a convergent continued fraction.
Unfortunately, in the contemporary literature at least, the
acceptance and credibility of the HD approach to polynomial $V(x)$
has been marred, in the quasi-variational context as well as in
perturbation theory, by a few unfortunate misunderstandings
concerning the related mathematics.

For a sketchy explanation of the latter point the reader should
consult the twenty years old Hautot's concise review
\cite{Hautot}. A more thorough account of the problem may be found
in my own long and unpublished dissertation \cite{disert}
including references to the misleading papers (responsible for the
unfortunate continuation of the whole misunderstanding up to these
days) as well as a detailed account and an exhaustive
specification of the domain of validity of the HD techniques in
Hermitian cases with polynomial potentials.

The present updated HD review is going to move one step further.
The reason is that in the light of some most recent semi-classical
analyses of anharmonic oscillators, another important motivation
for a renewal of interest in HD philosophy might be found in the
formal parallels between these two techniques, both having to deal
with complexified coordinates and both being based on the
infinite-series expansions of the energies $E$ and of the related
wave functions $\psi(x)$, respectively. It is also worth a
marginal note that in both of them a different role is played by
certain exponentially suppressed small contributions.

\section{${\cal PT}-$symmetric quantum mechanics}

Within the framework of the pure quantum physics, the idea of a
complexification of the couplings or coordinates is in fact
neither too popular nor formally welcome. A characteristic comment
has been made by the author of \cite{Simon} whose historical
remarks clarified the situation in the seventies when people
imagined that the convergence of perturbation series would imply
that the ``observables" may be well defined even when the
Hamiltonians themselves become manifestly non-Hermitian.

For this reason, perhaps, Carl Bender needed almost thirty years
more to return to the complexified models in the two pioneering
papers \cite{BB} where the two new versions of eq. (\ref{AHO})
have been studied with the purely imaginary couplings $b= i\beta$
and $d=i\delta$. The emphasis has been laid upon quite amazing an
observation that the spectrum remains real {\em in spite} of the
obvious fact that $H \neq H^\dagger$. In the light of the possible
applications of such an approach in Quantum Mechanics, the authors
characterized the underlying specific non-Hermiticity as a
``weakened Hermiticity" or ``${\cal PT}$ symmetry" of the
Hamiltonian. Indeed, in their most elementary models the operators
${\cal P}$ and ${\cal T}$ denoted the parity and the time reversal
(i.e., in effect, the Hermitian conjugation), respectively.

The methods they used were mainly perturbative and semi-classical.
In what follows, we intend to answer the question whether some
models of ${\cal PT}-$symmetric quantum mechanics could admit the
constructive solution of their Schr\"{o}dinger equations by the
techniques of the HD type. For the sake of definiteness and in a
way employing our recent results in \cite{tento}, we shall mainly
review the results concerning the above-mentioned quartic problem
 \be
  \left
(-\,\frac{d^2}{dx^2} + x^4+i\,\beta\,x^3+c\,x^2+i\,\delta\,x
 \right )
 \, \psi(x) =
E  \, \psi(x), \ \ \ \ \ x \in (-\infty,\infty)
 \label{SE}
 \ee
again, with the real values of $\beta,\,c$ and $\delta$ and,
therefore, with the manifestly non-Hermitian, ${\cal
PT}-$symmetric form of the Hamiltonian.

Before getting in more details, let us only add that during the
rapid development of the field (summarized, e.g., in the
proceedings of the recent Workshop which have to appear in the
October issue of Czechoslovak Journal of Physics this year), the
``parity" ${\cal P}$ should be understood, in a broader sense, as
an arbitrary (Hermitian and invertible) operator of a
pseudo-metric $\eta_{(pseudo)}$ in Hilbert space \cite{srni}.
Similarly, the symbol ${\cal T}$ should be understood as an
antilinear operator mediating the Hermitian conjugation of
$H$~\cite{BBJ}.

In the context of physical applications, the concept of the
``parity" itself becomes insufficient and must be complemented by
a ``quasi-parity" operator ${\cal Q}$ \cite{ptho} which makes the
scalar products (between eigenstates of $H \neq H^\dagger$)
positively definite \cite{pseudoQues}. This means that we must
perform a highly nontrivial transition from the non-singular and,
by assumption, indeterminate original pseudo-metric ${\cal P} =
\eta_{(pseudo)}=\eta_{(pseudo)}^\dagger$ to some other, positively
definite ``true"  metric ${\cal QP} = \eta_+ > 0$, the knowledge
of which allows us to call our Hamiltonians $H \neq H^\dagger$
``quasi-Hermitian" and ``physical" again (cf. the reviews
\cite{Dirac} for more details). In the other words, only the
change of the metric ${\cal P} \to {\cal QP}=\eta_{(quasi)}
=\eta_{(quasi)}^\dagger$ (where, at present, the name ``charge"
for ${\cal Q} \equiv {\cal C}$ is being preferred \cite{prl})
makes  the non-Hermitian ${\cal PT}-$symmetric models with real
spectrum fully compatible with the probabilistic interpretation
and other postulates of quantum mechanics in a way illustrated
very recently on the non-Hermitian square well \cite{sqw,132}.

\section{HD construction}

The wave functions $\psi(x)$ defined by eq. (\ref{SE}) are
analytic at all the complex $x \in l\!\!C$. In the asymptotic
region of the large $|x| \gg 1$ the semi-classical analysis
reveals that
 \be
 \psi^{(AHO)}(x) \sim \left \{
 \baa
 c_{phys}e^{-x^3/3}+
 c_{unphys}e^{+x^3/3}, \ \ &0 \leq |\,{\rm Im}\,(x)| \ll + \,{\rm
 Re}\,(x)/\sqrt{3}, \\
 d_{phys}e^{+x^3/3}+
 d_{unphys}e^{-x^3/3}, \ \ &0 \leq |\,{\rm Im}\,(x)| \ll - \,{\rm
 Re}\,(x)/\sqrt{3}.
 \ea
 \right .
 \label{estimates}
 \ee
%\subsection{Ansatz}
The HD method proceeds, traditionally, in an opposite direction
and employs the power-series formulae for $\psi(x)$ near the
origin. Thus, in the ${\cal PT}-$symmetric manner we write
 \be
\psi^{(ansatz)}(x)=e^{-sx^2}\,\sum_{n=0}^{\infty}\,h_n\,(ix)^n\,,
 \ \ \ \ \ x \in (-\infty,\infty)\,
 \label{ansatz}
 \ee
with a suitable (optional) value of $s$. The key reason may be
seen in the success of such a strategy in the harmonic-oscillator
case where such an ansatz converts the differential
Schr\"{o}dinger equation (\ref{SE}) into solvable recurrences.
Here, of course, the resulting recurrent rule
 \be
A_n\,h_{n+2} +C_n\,h_n + \delta \,h_{n-1} +
\theta\,h_{n-2}-\beta\,h_{n-3} + h_{n-4}=0 \label{recurrences}
 \ee
 \ben
A_n=(n+1)(n+2), \ \ \ C_n=4sn+2s-E, \ \ \ \theta = 4s^2-c
 \een
is not solvable in closed form. At all the parameters (including
also the unconstrained, variable energy $E$), it only defines the
coefficients $h_n$ as superpositions
 \be
  h_n=h_0\sigma_n+h_1\omega_n,
\ \ \ \ \ \ \ \ \ \ \sigma_0=\omega_1=1, \ \
 \sigma_1=\omega_0=0
 \ee
where all the three sequences $h_n$, $\sigma_n$ and $\omega_n$
satisfy the same recurrences with different initial conditions.
Still, we may write the latter two funtions of $n$ in a very
compact determinantal form
 \ben
  \sigma_{n+1}=(-1)^{n}\frac{\det
\Sigma_{n-1}}{n!(n+1)!}\,, \ \ \ \ \ \
 \omega_{n+1}=(-1)^{n}\frac{\det \Omega_{n-1}}{n!(n+1)!}\,, \ \ \
\ n = 1,2, \ldots\,
 \een
with $(m+1)-$dimensional matrices
 \be
  \Sigma_m=
  \left(
\begin{array}{ccccccc}
C_0&A_0&& & &&\\ \delta&0&A_1&& & &\\ \vdots&C_2&0&A_2&&& \\
1&\vdots&C_3&\ddots&\ddots&&
\\
0&-\beta&\ddots&\ddots&0&A_{m-2}&
\\
\vdots &1&\ddots&\delta&C_{m-1}&0&A_{m-1}
\\
 &&\ddots&\ldots&\delta&C_m&0
\\
\end{array} \right)
\label{4.8cc}
 \ee
(note: the second column of the matrix of the linear system
(\ref{recurrences}) is omitted here since $\sigma_1=0$) and
 \be
  \Omega_m=
  \left(
\begin{array}{ccccccc}
0&A_0&& & &&\\ C_1&0&A_1&& & &\\
\delta&C_2&0&A_2&&& \\ \vdots&\delta&C_3&\ddots&\ddots&&
\\
1&\vdots&\delta&\ddots&0&A_{m-2}&
\\
 &\ddots&\vdots&\ddots&C_{m-1}&0&A_{m-1}
\\
 &&1&\ldots&\delta&C_m&0
\\
\end{array} \right)
\label{4.8bb}
 \ee
(now the first column is omitted in the light of our choice of $
\omega_0=0$).
%\subsection{Free parameters}

It remains for us to specify the norm $\rho=\sqrt{h_0^2+h_1^2}$,
the ratio $h_1/h_0 \equiv \tan \zeta$ and the physical energy $E$
which enters the $(m+1)-$dimensional matrices (\ref{4.8cc}) and
(\ref{4.8bb}). In the other words, our wave functions $\psi(x)$
must be made ``physical" via the standard boundary conditions
 \be
 \psi^{(ansatz)}(X_R) = 0= \psi^{(ansatz)}(-X_L)\,, \ \ \ \ \ \
 X_R \gg 1, \ \  X_L \gg 1\,.
 \label{bc}
 \ee
These conditions play, obviously, the role of an implicit
definition of the above free parameters. Nevertheless, our
ambitions are higher and in the HD approach people usually try to
replace eq. (\ref{bc}) by a quasi-variational finite-matrix
truncation of eq. (\ref{recurrences}) \cite{Hautot}, i.e.,
equivalently, by innovated difference-equation boundary conditions
 \be
 h_N=h_{N+1} = 0\,, \ \ \ \ \ \
 N \gg 1\,.
 \label{bcd}
 \ee
with a much simpler numerical implementation. Of course, such a
replacement of eq. (\ref{bc}) by eq. (\ref{bcd}) is not always
well founded \cite{Hautot} so that in each particular construction
its validity must be based on a rigorous mathematical proof, in a
way illustrated in what follows.

\section{The proof}

\subsection{The $s-$dependence of the asymptotics of $h_n$}

In the first step of the proof, the recurrences
(\ref{recurrences}) should be read as a linear difference equation
of the sixth order. The sextuplet of its independent asymptotic
solutions $h_n$ may be found by the standard techniques. Thus, the
leading-order solution may be extracted from eq.
(\ref{recurrences}) by its reduction to its two-term dominant form
of relation between $h_{n+2}$ and $h_{n-4}$. Thus, we put
 \be
h_n(p)=\frac{\lambda^n(p)\,g_n(p)}{(3^{1/3})^n\,\Gamma(1+n/3)}
 \label{coeff}
 \ee
where the integer $p = 1, 2, \ldots, 6$ numbers the six
independent solutions and where $\lambda(p) =\exp [i(2p-1)\pi/6]$.
The new coefficients $g_n=g_n(p)$ vary more slowly with $n$. Thus,
our first conclusion is that at the large indices $n$, all
solutions decrease as $h_n \sim {\cal O}(n^{-n/3})$ at least. This
means that in a way confirming our expectations the radius of
convergence of our Taylor series (\ref{ansatz}) is always
infinite.

On this level of precision the size of our six independent
solutions remains asymptotically the same. In order to remove this
degeneracy we amend equation (\ref{recurrences}) and having
temporarily dropped the $p$'s we get
 \be
 g_{n+2} - g_{n-4} = \frac{4s\lambda^4}{n^{1/3}}\,g_n
 - \frac{\beta\,\lambda}{n^{1/3}}\,g_{n-3} + {\cal O}
 \left ( \frac{g_n}{n^{2/3}} \right ) .
 \label{cond}
 \ee
The smallness of  $1/n^{1/3}$ in the asymptotic region of $ n \gg
1$ enables us to infer that
 \be
 g_n= e^{\gamma \,n^{2/3}+{\cal O}(n^{1/3})}, \ \ \ \ \gamma =
 \gamma(p)=
  s \lambda^4(p)- \beta\,\lambda(p)/4,
\label{second}
 \ee
i.e.,
 \ben
{\rm Re}\ \gamma (1)={\rm Re}\ \gamma (6) = -\frac{\sqrt{3}}{8}
\beta - \frac{s}{2}, \ \ \ \ \
 {\rm Re}\ \gamma (2)={\rm Re}\ \gamma
(5) = s,
 \een
 \be
  {\rm Re}\ \gamma (3)={\rm Re}\ \gamma (4) =
\frac{\sqrt{3}}{8} \beta -
 \frac{s}{2}.
 \label{estim}
  \ee
We may conclude that whenever we satisfy the condition
 \be
 s > \frac{| \beta |}{4\sqrt{3}}
\label{constr}
 \ee
the general, six-parametric form of the Taylor coefficients
 \be
  h_n=\sum _{p=1}^6 G_p\,h_n(p)
  \label{ress}
  \ee
will be equivalent to the two-term formula in the leading order,
 \be
 h_n=G_2\,h_n(2)+G_5\,h_n(5), \ \ \ \ \ \ \ n \gg 1,
\label{gensol}
  \ee
i.e., only its two dominant components remain asymptotically
relevant while, in the leading order, we may simply put $
G_1=G_3=G_4=G_6=0$ in eq. (\ref{ress}) at $n \gg 1$.

The same argument implies that for
 \be
 s < \frac{| \beta |}{4\sqrt{3}}
\label{constrol}
 \ee
we get
 \be
 h_n= \left \{
 \ba
 G_3\,h_n(3)+G_4\,h_n(4), \ \ \ \ \ \ \ \beta > 0, \ n \gg 1,\\
 G_1\,h_n(1)+G_6\,h_n(6), \ \ \ \ \ \ \ \beta < 0, \  n \gg 1.
 \ea
 \right .
\label{gensoll}
  \ee
while the degeneracy of more than two solutions survives at
$\beta=0$ or at $s = \frac{| \beta |}{4\sqrt{3}}>0$.

\subsection{Single-term dominant exponentials in $\psi^{(ansatz)}(x)$ }

We know that the shape of the function $\psi^{(ansatz)}(x)$ is
determined by the energy $E$ and by the not yet fully specified
choice of the two coefficients $h_0$ and $h_1$. In the language of
the preceding section, each choice of the energy $E$ and of the
initial $h_0$ and $h_1$ will generate a different, $x-$ and
$n-$independent pair of the coefficients $G_{2}$ and $G_{5}$ in
$h_n$ at the sufficiently large $s$ (the discussion of the smaller
$s$ will be omitted for the sake of brevity).

As long as our aim is a completion of the proof of the validity of
the replacement of the standard boundary conditions (\ref{bc})
(containing infinite sums) by the more natural HD truncation of
recurrences due to eq. (\ref{bcd}), we just have to parallel the
Hermitian considerations of ref. \cite{Classif}. Firstly, we
remind the reader that at an arbitrary finite precision we always
have $E \neq E(physical)$. This means that our infinite series
$\psi^{(ansatz)}(x)$ as defined by equation (\ref{ansatz}) will
{\em always} exhibit an exponential asymptotic growth in a certain
wedge-shaped vicinity of the real line \cite{BB}.

The main point of the HD proof is that due to the standard
oscillation theorems the last (left as well as right) nodal zeros
will move {\em towards} their (left and right) infinities with the
decrease of $E> E(physical)$ (say, at a fixed pair $h_{0,1}$). In
the language of coordinates this means that the values of
$\psi^{(ansatz)}(x)$ (both at some $x>X_{R}$ and at another
$x<X_{L}$) will change sign for $E$ somewhere in between the
(smaller) $ E(physical)$ and (larger) $E(numerical)$. Such a
simultaneous change of the sign of $\psi^{(ansatz)}(x)$ at any
sufficiently large absolute value of the coordinate $|x| \gg 1$
should be understood as resulting from our appropriate asymptotic
estimate of $\psi^{(ansatz)}(x)$.

Although this sounds like a paradox, this estimate will {\em
always} lead to an asymptotic growth (since $E \neq E_{physical}$
with probability one) so that it cannot be changed by the (key)
omission of any {\em finite} number $N$ of the exponentially small
components $\sim \exp [-sx^2 + {\cal O}(\ln x)]$. They may be
safely ignored as irrelevant. This means that we may choose $N \gg
1$ so that just the asymptotically dominant coefficients will play
the role. Inserting (\ref{coeff}) and (\ref{gensol}) in
$\psi^{(ansatz)}(x) \sim \exp(-sx^2)\,\sum_{n=N+1}^{\infty}\, h_n
\,(ix)^n$ we finally get
 \ben
 \psi^{(ansatz)}(x) \sim
e^{-sx^2}\,\sum_{n=N+1}^{\infty}\,
\frac{G_2\lambda^n(2)\,g_n(2)+G_5\lambda^n(5)\,g_n(5)
}{(3^{1/3})^n\,\Gamma(1+n/3)} \,(ix)^n, \ \ \ \ \ \ |x| \gg 1.
 \een
The validity of this formula is a strict consequence of the
specific constraint (\ref{constr}) imposed (say, from now on) upon
the admissible quasi-variational parameter $s$.

Once we split $ \psi^{(ansatz)}(x) =\psi^{(ansatz)}(G_2,G_5,x)$ in
its two components
 \ben
 \psi^{(ansatz)}(G_2,0,x) \sim G_2\,
e^{-sx^2}\,\sum_{n=N+1}^{\infty}\, \frac{(-x)^n\, \exp \left [
\gamma(2)\,n^{2/3} + {\cal O} (n^{1/3})
 \right ] }{(3^{1/3})^n\,\Gamma(1+n/3)},
 \label{bensa}
 \een
 \ben
 \psi^{(ansatz)}(0,G_5,x) \sim G_5\,
e^{-sx^2}\,\sum_{n=N+1}^{\infty}\, \frac{x^n\, \exp \left [
\gamma(5)\,n^{2/3} + {\cal O} (n^{1/3})
 \right ] }{(3^{1/3})^n\,\Gamma(1+n/3)},
 \label{cens}
 \een
we may apply the rule $e^z \sim (1+z/t)^t, \ t \gg 1$ in the error
term and get
 \ben
  \frac{ \psi^{(ansatz)}(G_2,0,-y)}
{\exp(-sy^2)}
  \sim G_2\,
\sum_{n=N+1}^{\infty} \frac{1}{(3^{1/3})^n\,\Gamma(1+n/3)}
 \left \{y \cdot
 \left [1+{\cal O}
 \left(\frac{1}{N^{1/3}}
 \right )
 \right ]
 \right \}^n
 \label{bsa}
 \een
and
 \ben
 \frac{
 \psi^{(ansatz)}(0,G_5,y)}
{\exp(-sy^2)}
  \sim G_5\,
\sum_{n=N+1}^{\infty} \frac{1}{(3^{1/3})^n\,\Gamma(1+n/3)}
 \left \{y \cdot
 \left [1+{\cal O}
 \left(\frac{1}{N^{1/3}}
 \right )
 \right ]
 \right \}^n.
 \label{cs}
 \een
This is valid at all the large arguments $y$. Along the {\em
positive} semi-axis $y \gg 1$, both the right-hand-side summands
are real and positive. They sum up to the same function
$\exp[y^3/3+ {\cal O}(y^2)]$. This is a consequence of the
approximation of the sum by an integral and its subsequent
evaluation by means of the saddle-point method. The same trick was
used by Hautot, in similar context, for the ${\cal P}-$symmetric
and Hermitian anharmonic oscillators \cite{Hautot}.

In contrast to the Hautot's resulting one-term estimates of
$\psi$, the present asymmetric, ${\cal PT}-$invariant construction
leads to the more general two-term asymptotic estimate
 \ben
 \psi^{(ansatz)}(G_2,G_5,x) \sim G_2\,
\exp[-x^3/3+ {\cal O}(x^2)]+
 G_5\, \exp[x^3/3+ {\cal O}(x^2)], \ \ \ \ |x| \gg 1.
 \label{finsa}
 \een
As long as we deal with the holomorphic function of $x$, this
estimate may be analytically continued off the real axis of $x$.
Near both the ends of the real line and within the asymptotic
wedges $|{\rm Im}\ x|/|{\rm Re}\ x| < \tan \pi/6$ we simply have
the rules
 \be
 \psi^{(ansatz)}(G_2,G_5,x) \sim G_2\,
\exp[-x^3/3+ {\cal O}(x^2)], \ \ \ \ {\rm Re}\ x <-X_L \ll - 1
 \label{leva}
 \ee
and
  \be
 \psi^{(ansatz)}(G_2,G_5,x) \sim G_5\,
\exp[x^3/3+ {\cal O}(x^2)], \ \ \ \ 1 \ll X_R <{\rm Re}\ x .
 \label{prava}
 \ee
They are fully compatible with our {\it a priori} expectations and
represent in fact the main step towards our forthcoming completion
of the rigorous proof of the validity of the HD matrix truncation
(\ref{bcd}).

\subsection{The changes of sign of $\psi^{(ansatz)}(x)$ at $|x| \gg
1$ }

Our complex differential Schr\"{o}dinger equation (\ref{SE})
becomes asymptotically real, in the leading-order approximation at
least. In a suitable normalization the wave functions
$\psi^{(ansatz)}(x)$ may be made asymptotically real as well. Near
infinity they will obey the standard Sturm Liouville oscillation
theorems. As we explained, after a small decrease of the tentative
energy parameter $E> E(physical)$ the asymptotic nodal zero
$X_{R}$ or $-X_{L}$ originating in one of our boundary conditions
(\ref{bc}) will move towards infinity. This may be re-read as a
doublet of conditions
\be
G_2= G_2(E_0,\zeta_0)=0, \ \ \ \ \ \ G_5=G_5(E_0,\zeta_0) = 0
\label{newbc}
 \ee
using an assumption that  $\zeta_0 \approx \zeta(physical)$ in the
suitable parametrization of $h_0=\rho\,\cos \zeta$ and
$h_1=\rho\,\sin \zeta$ with $\zeta \in (0, 2\pi)$ at a convenient
normalization $\rho=1$.

In the limit $N \to \infty$ the two requirements (\ref{newbc}) may
be interpreted as equivalent to the truncation recipe (\ref{bcd}).
Indeed, at a fixed $N \gg 1$ we may re-scale
 \ben f_p=G_p
\frac{\lambda^N(p)\,\exp [\gamma(p) N^{2/3}]}
{(3^{1/3})^N\,\Gamma(1+N/3)}, \ \ \ \ \ \ p = 2,5
 \label{abbrevici}
 \een
where $f_p(E,\zeta_0) \approx F_p \cdot (E-E_0)$. This enables us
to write
 \ben \ba h_N \approx
(F_2+F_5)(E-E_0)+{\cal O}[(E-E_0)^2],\\ (N+3)^{1/3}h_{N+1} \approx
[F_2\lambda(2)+F_5\lambda(5)](E-E_0) +{\cal O}[(E-E_0)^2] \ea
 \een
due to equation (\ref{coeff}). We see that the two functions
$G_2,\,G_5$ are connected with the Taylor coefficients
$h_N=h_N(E_0,\zeta_0)$ and $h_{N+1}=h_{N+1}(E_0,\zeta_0)$ near the
physical $E_0$ and $\zeta_0$ by an easily invertible regular
mapping.  This means that the implicit algebraic boundary
conditions (\ref{newbc}) are strictly equivalent to the fully
explicit requirements
 \be
  h_N(E_0,\zeta_0) = 0, \ \ \ \ \ \ \ \ \
h_{N+1}(E_0,\zeta_0) = 0, \ \ \ \ \ \ N \gg 1. \label{nec}
 \ee
This completes our proof.

\section{Discussion}

\subsection{Truncated secular Hill determinants}

By construction, our present result may be re-read as a
demonstration that the intuitive quasi-variational square-matrix
truncation of our recurrences represents in fact a mathematically
well founded approximation recipe. The evaluation of both the
energies and wave functions may be started at any approximative
cut-off $N<\infty$ and the solution of the linear algebraic
problem
 \be
  \left(
\begin{array}{ccccccc}
C_0&0&A_0&& & &\\ \delta&C_1&0&A_1&& & \\
\theta&\delta&\ddots&\ddots&\ddots&& \\ -\beta&\theta&\ddots&&&&
\\
 1&-\beta&\ddots&&&\ddots&A_{N-3}
\\
 &\ddots&\ddots&\ddots&\ddots&\ddots&0
\\
 &&1&-\beta&\theta&\delta&C_{N-1}
\\
\end{array} \right) \left( \ba
h_0\\ h_1\\ h_2\\ \ldots\\h_{N-3}\\ h_{N-2}\\ h_{N-1} \ea \right)
= 0 \label{4.8}
 \ee
must only be completed by the limiting transition $N \to \infty$.
As long as the energy only enters the main diagonal,
$C_n=4sn+2s-E$, we may determine all the approximate low-lying
spectrum by the routine $N \times N-$dimensional asymmetric-matrix
diagonalization. With $s=2$ and $a=c=\beta=\delta=1$, Table~1
illustrates the practical implementation as well as the highly
satisfactory rate of convergence of the numerical HD algorithm.

\subsection{Outlook}

We have re-confirmed that the HD methods bridge a gap between the
brute-force variational algorithms and sophisticated analytic
semi-classical constructions. In the present context, we
emphasized that the HD techniques may be combined not only with
the complexification of the variables (characteristic for the
latter approach) but also with the requirements of a numerical
efficiency (pursued, usually, in the former setting).

In the conclusion we may note that another link connects also the
HD and perturbative techniques. In the latter, semi-analytic type
of considerations, an indispensable role is played by asymptotic
expansions, say, of the power-series form
 \be
 Z(\lambda) = \sum_{k=0}^M \lambda^k z_k + R_M(\lambda)
 \label{sum}
 \ee
where $\lambda$ represents a - presumably, ``sufficiently" small -
coupling constant while the typical example of the observable
$Z(\lambda)$ is a bound-state energy. In the present HD setting we
employed a similar ansatz (\ref{sum}) where the variable $\lambda$
denoted the coordinate $x$ while the function $Z(\lambda)$
represented the wave function.

Currently, a numerical credibility of the similar perturbative
(and also semi-classical) constructions is being enhanced via an
improved treatment of the error term $R_M(\lambda)$. In general,
this does not seem to be the case in the present HD context. The
reason is that although for the bound states the exponentially
growing right-hand sum must be identically zero, we are still not
interested in the explicit evaluation of the exponentially
suppressed remainder $R_M(\lambda)$ (which would be hopeless) but
merely in the bracketing property of the wave-function sum
$\sum_{k=0}^M \lambda^k z_k \neq  0$ evaluated {\em slightly below
and slightly above} the actual physical energy.

In this sense, we may summarize that the HD ``trick" is different
and enables us to pay a more detailed attention to the new
perspectives opened by the consequent complexification of
$\lambda$. Here, such a perspective enabled us to extend the
standard mathematical background of the Hill-determinant
constructions of bound states. We have seen that the old
principles stay at work, making use of the same mechanism as in
the standard Hermitian cases, viz, the constructively employed
asymptotic cancellation of the growing exponentials in the
exponentially decreasing wave functions $\psi(x)$.

\section*{Acknowledgements}
Work partially supported by RIMS, Kyoto, and by the AS CR grant
Nr.  A 1048302.

\section*{Table}

%\newpage

Table 1. The convergence of HD energies.

$$
   \begin{array}{||cc|lllll||}
\hline \hline &&\multicolumn{5}{|c||}{{ energies\ } E_n}\\ \hline
 &n&0&1&2&3&4
 \\ N&&&&&&\\ \hline 15&& 1.69347&
    5.106&
    9.152&
    13.043&
    17.82\\
20&& 1.691638&
    5.1256&
    9.2800&
    14.050&
    19.24\\
25& &1.691579&
    5.12344&
    9.2581&
    13.869&
    18.79\\
30&& 1.691590&
    5.12361&
    9.2617&
    13.883&
    18.89\\
35& &1.691590&
    5.12358&
    9.2615&
    13.879&
    18.88
    \\
\hline
\hline
\ea
$$

% \newpage

\end{document}